\documentclass[twocolumn,preprintnumbers,amsmath,amssymb,floatfix]{revtex4}
\usepackage{graphicx}
\usepackage{amsmath}
\usepackage{bm}
\usepackage{mathrsfs}
\usepackage{color}
\usepackage{slashed}
\usepackage{dcolumn}

\newcommand{\bald}[1]{{\bf #1}}
\newcommand{\baldh}[1]{\hat{\bf #1}}

\newcommand{\eqf}[1]{\begin{equation}\begin{split}#1\end{split}\end{equation}}

\begin{document}

\title{Jet momentum balance independent of shear viscosity}

\author{R. B. Neufeld}
\address{Los Alamos National Laboratory, Theoretical Division, MS B238, Los Alamos, NM 87545, U.S.A.}
\email{neufeld@lanl.gov}

\date{\today}

\begin{abstract}
Jet momentum balance measurements, such as those recently performed by the CMS collaboration, provide an opportunity to quantify the energy transferred from a parton shower to the underlying medium in heavy-ion collisions.  Specifically, I argue that the Cooper-Frye freezeout distribution associated with the energy and momentum deposited by the parton shower is controlled to a significant extent by the distribution of the underlying bulk matter and independent of the details of how deposited energy is redistributed in the medium, which is largely determined by transport coefficients such as shear viscosity.  Thus by matching the distribution of momentum associated with the secondary jet in such measurements to the thermal distribution of the underlying medium, one can obtain a model independent estimate on the amount of parton shower energy deposited.
\end{abstract}

\maketitle

\section{Introduction}

The suppression in energetic leading particle production rates in relativistic heavy-ion collisions relative to that expected from a naive superposition of nucleon-nucleon collisions has long been regarded as a signature of the formation of the quark gluon plasma (QGP) \cite{Bjorken:1982tu}.  The observation of this suppression in high $p_T$ measurements is one of the most striking results from the heavy-ion program at the Relativistic Heavy-Ion Collider (RHIC) \cite{Adcox:2001jp} and now at the Large Hadron Collider (LHC) \cite{LHCII}.  This phenomena, often referred to as ``jet quenching'' \cite{Gyulassy:2003mc}, is largely reflective of the interaction of energetic, or fast, partons with the QGP and is a useful probe of the hot and dense matter formed in relativistic heavy-ion collisions.

Jet observables are more differential than leading particle observables and have the potential to probe the QGP and partonic energy loss dynamics in new and powerful ways.  The definition of a jet depends on parameters such as the jet cone radius and minimum energy acceptance cuts.  The freedom one has in choosing these parameters can be exploited to gain new insights into the physics of the jet-medium interaction.  A fast parton evolves into an in-medium parton shower through medium-induced radiative processes \cite{jet1,jet2}.  The parton shower is further modified through collisional energy losses to the underlying medium \cite{ev1,ev2,ev3} which serve to alter the spectrum of radiated gluons.  Jet observables have been proposed as a channel through which to constrain the final shape and energy distribution associated with this shower \cite{jetty,Renk:2009hv}.  For example, the variation of the jet nuclear modification factor, $R_{AA}$, with the jet cone radius may be sensitive to the final angular distribution of emitted radiation.

Parton showers can be substantially modified through collisional energy losses - or  {\it energy deposition} - to the underlying medium.  The dynamics of the parton shower energy deposition is reflective of the properties of the QGP and is also sensitive to the distribution of medium induced radiation - such as the gluon multiplicity and angle of emission.  Knowing the amount of energy deposited by a parton shower is thus not only important for understanding jet measurements but also provides another window into the properties of medium induced radiation.  

In this paper, I show that jet momentum balance measurements, such as those recently performed by the CMS collaboration \cite{LHCIII} at the LHC, provide an opportunity to quantify parton shower energy deposition in heavy-ion collisions.  This is true because, as will be argued for and shown on a specific example below, the Cooper-Frye freezeout spectrum associated with the energy and momentum deposited by the parton shower is largely controlled by the distribution of the underlying bulk matter and independent of the details of how deposited energy is redistributed in the medium.  Thus by matching the distribution of momentum associated with the secondary jet in such measurements to the thermal distribution of the underlying medium, one can obtain a model independent estimate on the amount of parton shower energy and momentum deposited.  

In the case of dijets this model independent estimate is for the net momentum deposited by the two (or more) parton showers.  To estimate the energy and momentum deposited by a single parton shower in dijet measurements requires some model estimate on relative pathlength and parent parton species.  However, for jet momentum balance measurements performed on jets tagged with electroweak bosons, one can obtain a model independent estimate of single parton shower energy deposition.

It is non-trivial that the freezeout spectrum associated with parton shower energy deposition may be largely independent of the details of how deposited energy is redistributed in the QGP.  When energy and momentum are deposited, the disturbance is initially localized around the axis of propagation of the parton shower, but the medium acts to redistribute the energy and momentum to larger angles and softer scales.  The details of how this redistribution occurs, or the medium response to the parton shower, depends to a large extent on the shear viscosity and speed of sound of the bulk medium.  In particular, a more viscous medium tends to dissipate the energy to softer scales more quickly than a less viscous medium.  One might well expect the freezeout spectrum associated with parton shower energy deposition to depend sensitively on the shear viscosity of the medium.  However, as will be argued in what follows, this is not the case.  Rather, the final spectrum associated with the parton shower energy deposition is constrained largely by the distribution of the underlying medium, which in turn can be used to quantify the amount of energy deposited.

The paper is laid out as follows.  In section \ref{II} the CMS dijet momentum balance measurement is discussed.  I argue that the results of the dijet momentum balance measurement have a natural interpretation in terms of parton shower energy deposition.  I further argue why the freezeout spectrum associated with the parton shower energy deposition might depend only weakly on the details of how the medium redistributes that energy - which is largely determined by the speed of sound and shear viscosity.  I then demonstrate this on the specific example of a parton shower in linearized hydrodynamics.  This simplified example provides a nice testing ground to show the general principle which may hold for more complicated examples as well.  In section \ref{III} the CMS jet momentum balance measurement results are examined and compared to data on bulk particle production at LHC energies.  From this comparison, I estimate that in dijet events with asymmetry of 0.4, the secondary jet deposits about 30-40 GeV of energy.  Section \ref{IIII} concludes with summary remarks.

\section{Cooper Frye Freezeout from a Parton shower in hydrodynamics}\label{II}

This section begins with a discussion of  the dijet momentum balance measurement performed by CMS \cite{LHCIII}.  In their analysis, CMS looked at jet production in PbPb collisions at a nucleon-nucleon center-of-mass energy of $2.76$ TeV.  They considered events with a leading jet of ${p_T}_1 > 120$ GeV and secondary jet of ${p_T}_2 > 50$ GeV, with no explicit requirement made either on the presence or absence of a third jet in the event.  Using an iterative cone algorithm jets were reconstructed with radius of 0.5 and minimum $p_T$ acceptance of 1.0 GeV.

CMS observed a significant dijet asymmetry, $A_j \equiv ({p_T}_1 - {p_T}_2)/({p_T}_1 + {p_T}_2)$, in PbPb collisions relative to pp.  Such an enhanced asymmetry is reflective of medium modification of parton showers in which energy and momentum are transferred out of the jet reconstruction phase space parameters.  One way to help quantify the distribution of this missing momentum is through the overall momentum balance of the dijet event.  In this analysis, the $p_T$ tracks of an event are projected onto the axis of the leading jet and summed over to explicitly account for overall momentum conservation in the event.  When averaged over many events, CMS found that a surplus of tracks with momentum projection greater than $8$ GeV in the hemisphere of the leading jet were balanced by those with momentum projection less than $8$ GeV in the opposite hemisphere, the majority of which were between $0.5 - 2.0$ GeV for central events.  The distribution of the momentum balance in the hemisphere opposite the leading jet was much softer than that predicted by PYTHIA + HYDJET.

The physical picture that emerges from the CMS dijet momentum balance measurement is that the secondary jet, which on average must traverse more of the hot and dense medium formed in heavy-ion collisions than the leading jet, is modified by interactions with the medium in such a way that the energy and momentum lost by the jet is transferred to relatively soft scales.  I here argue that the CMS results on dijet momentum balance have a natural interpretation in terms of the transfer of energy and momentum from a medium induced parton shower to the underlying medium which then carries that energy and momentum to larger angles and softer scales through transport processes.  

As will be demonstrated in the rest of this section, the amount of this parton shower energy deposition can be quantified by matching the distribution of momentum associated with the secondary jet in such measurements to the thermal distribution of the underlying medium.  The effect of parton shower energy and momentum deposition is to generate a disturbance in the underlying medium which shows up in the momentum balance of a heavy-ion collision.  Provided that this disturbance is not too strong, it is not unreasonable to think that one can obtain an estimate on parton shower energy deposition by matching the distribution of momentum associated with energy deposition to the thermal distribution of the underlying medium.  However, what is not obvious is how close this matching is to the actual energy deposition, or how sensitively the results depend on the details of how the medium redistributes that energy.  Further, one has the added subtlety that jet momentum balance measures momentum, which vanishes on the whole for the underlying medium (in the absence of jets).  So it is necessary to define what is meant by matching the distribution associated with parton shower energy deposition to that of the unperturbed medium.  This section aims to give some clarity to these details.

\subsection{The Parton Shower Spectrum from Cooper Frye}

Consider the Cooper-Frye (CF) \cite{Cooper:1974mv} freezeout prescription for converting energy and momentum into a final state particle distribution $dN$:
\eqf{\label{cfbasic}
E\frac{d N}{d^3 p} = \int \frac{d^3 \sigma_\mu \, p^\mu }{(2\pi)^3}\,f(p,x),
}  
where $E$ and $\bald{p}$ are the energy and momentum of particles in the final state and $f(p,x)$ is the distribution function just before freezeout.  In Equation (\ref{cfbasic}) $d^3 \sigma_\mu$ is the normal vector to the freezeout hypersurface and any degeneracy factors or summation over final state particles has been suppressed.  For the purposes considered here, the important feature of (\ref{cfbasic}) is that energy and momentum are explicitly conserved.  This is particularly important when considering the example of a parton shower in hydrodynamics, where it is necessary to track the energy and momentum deposited by the parton shower from the time of deposition all the way through to the final particle spectrum.

In this section the orientation of $d^3 \sigma_\mu$ will be determined by the local fluid velocity of the medium.  This choice corresponds to freezing out at a constant proper time and has the advantage that the local flow velocity takes an especially transparent form: $\bald{u} \equiv \bald{g}/ \epsilon$, where $\bald{g}$ is the momentum density of the medium and $\epsilon$ is the energy density \cite{Ivanov:2008zi}.  In the implementation that follows I will ignore differences between the proper time and the global time as these differences are of higher order in the solution to linearized hydrodynamics.  Thus the time of freezeout in the proper frame will be the same as in the frame of computation.  Additionally, final state particles will be taken as massless and described by a Boltzmann distribution for simplicity.  Viscous corrections to the distribution function will not be considered.  One expects these corrections may be small in the momentum range of interest, however this must be quantified.  In this case, one has
\eqf{\label{cfspecific}
E\frac{d N}{d^3 p} = \int \frac{d^3 x \, p \,(1-\bald{u}\cdot \baldh{p} )}{(2\pi)^3}\, \exp \left[-\frac{p}{T}\gamma (1-\bald{u}\cdot \baldh{p} )\right]
}
where $\bald{u}$ is the local fluid velocity, $\gamma = 1/\sqrt{1-u^2}$ and $T$ is the temperature at freezeout.  When speaking of the distribution of the unperturbed, or underlying thermal, medium it is here meant
\eqf{\label{underlie}
\int d^3 p \, E\frac{d N}{d^3 p} \text{     ,    } p_1 < |p| < p_2
}
where $E\frac{d N}{d^3 p}$ is obtained in the case of no parton shower energy deposition.  The differential form of (\ref{underlie}) will also be used later, which is obtained by not performing the $d |p|$ integration.  In the language of a heavy-ion collision, if one ignores mass effects and focuses on the region of mid-rapidity, then the underlying particle distribution is
\eqf{\label{experiment}
\int d p_T \, p_T \frac{d N}{d p_T} \text{     ,    } p_1 < p_T < p_2
}
which can be evaluated directly from experimental data.  Equation (\ref{experiment}) will be put into use in section \ref{III}.

To analyze the momentum balance associated with a parton shower one needs the total momentum.  This quantity, which evaluates to zero for a medium without parton shower energy deposition, is obtained from (\ref{cfspecific}) as
\eqf{\label{momentum}
\bald{P} &= \int d^3 p \, \bald{p} \,\frac{d N}{d^3 p} \\
&= \int \frac{d^3 p\,d^3 x \, \bald{p} \,(1-\bald{u}\cdot \baldh{p} )}{(2\pi)^3}\, \exp \left[-\frac{p}{T}\gamma (1-\bald{u}\cdot \baldh{p} )\right]. 
}
The reader can verify that (\ref{momentum}) conserves momentum in the sense that $\bald{u} \equiv \bald{g}/ \epsilon$, where as mentioned previously $\bald{g}$ is the momentum density and $\epsilon$ is the energy density.  The total momentum within some interval, $p_1 < |p_i| < p_2$, is obtained by limiting the integration in (\ref{momentum}) to the appropriate region (the component $i$ will be determined by the parton shower direction of propagation).  This is what is meant by the distribution of momentum associated with parton shower energy deposition, and will be referred to as the {\it momentum deposition spectrum} in what follows.  Again, one can also use the differential form of (\ref{momentum}) by not performing the $d |p_i|$ integration.  Note that the momentum deposition spectrum does not contain all of the energy and momentum 'lost' by a parton shower, but only that fraction that goes into the underlying thermal medium.

In the limit that the effect of the parton shower on the underlying medium is small - small meaning that on average the flow velocity and temperature fluctuations can be linearized - the energy and momentum perturbations generated by the parton shower decouple from the momentum dependence of the final particle distribution.  For a static background this ensures that the momentum deposition spectrum is completely controlled by the distribution of the underlying medium - although it is not immediately obvious in precisely what functional form.    However, it turns out that the functional form is exactly the same as that of the underlying medium in this limit.

To see this explicitly, consider the static background approximation in which parton shower energy and momentum deposition induces a small flow velocity in the direction of its propagation, which is taken to be in the $\hat{x}$ direction.  The momentum within some interval, $p_1 < p_x < p_2$, is found from (\ref{cfspecific}) to be
\eqf{
P_x &= \int \frac{d^3 x \,u(x)}{(2\pi)^2} \int_{-1}^{1} d v \int_{p_1/|v|}^{p_2/|v|} d p \, p^3\,v^2 \,\frac{p-T}{T} e^{-\frac{p}{T}}, 
}
where any temperature variation generated by the parton shower is of higher order in the linearization.  One can further proceed with a change in variables $m = p/(|v| T)$ followed by integrating by parts to find
\eqf{\label{simply}
P_x &= \int  d^3 x \,g_x(x) \, \frac{1}{6}\int_{p_1/T}^{p_2/T} d m \, m^3\, e^{-m} 
}
where $g_x(x)$ is the local momentum density, as mentioned above.  It is straightforward to check that this has the exact same form as the particle distribution of the unperturbed medium, normalized to the total momentum deposited by the parton shower.

The result of (\ref{simply}) suggests what was stated earlier: that the momentum deposition spectrum generated by a parton shower is largely controlled by the distribution of the underlying bulk matter.  The result (\ref{simply}) no longer holds if the linearized approximation breaks down, even in the case of a static background.  However, for a realistic amount of energy deposition, the linearized approximation is relatively accurate when integrating over the entire volume - provided the energy deposition is not too localized.

\subsection{A Specific Example - Parton Shower in Linearized Hydrodynamics}

The details of the implementation of linearized hydrodynamics coupled to a parton shower source term are discussed in detail in \cite{Neufeld:2010tz}.  In this section I will sketch some of the qualitative aspects of this implementation for pedagogical purposes, leaving the interested reader to refer to the above work for more detail.
In the linearized approximation one assumes a static background.  Although the static approximation is not realistic for a heavy-ion collision, two of the important features, energy and momentum conservation and correct scaling with shear viscosity, remain intact.  In this way one hopes to obtain qualitative information about how the momentum deposition spectrum relates to the distribution of the underlying medium in a heavy-ion collision.

Of course, a more realistic treatment would be to incorporate parton shower energy deposition in a full viscous hydrodynamic simulation of a heavy-ion collision.  Then one could evaluate the momentum deposition spectrum and compare to the distribution in events without energy deposition.  However, in order to make a complete comparison, it is necessary to vary the dependence on the shear viscosity as well as consider other possibly important details to ensure these do not significantly alter the results.  The linearized approximation provides a convenient way to consider these variations in this first study.  

It is important to distinguish between the linearization performed in the solution to hydrodynamics in what follows and the linearization performed above in the CF equation.  The linearization done above was to illustrate what happens in a certain limit, but is not adopted in what follows.  Rather, the results obtained from the hydrodynamic solution are kept in their full functional form when evaluating the final spectrum.  This is consistent even within the regime of validity of linearized hydrodynamics because the expansion of the CF equation involved both the perturbation due to the parton shower and $p/T$, which need not be small.  Additionally, even if one applies linearized hydrodynamics in a regime where the linearization breaks down, the energy and momentum are still conserved, so there is at least a qualitative description of the underlying dynamics.

In any hydrodynamic simulation coupled to a parton shower it is necessary to specify the source term which represents the flow of energy and momentum between parton shower and medium.  The simplest form for the source which conserves energy and momentum is to treat the parton shower as a point source with a weight given by the energy loss rate to the medium:
\eqf{\label{simplesource}
J^\nu = \frac{dE}{dt}(t) \, u^\nu\delta(\bald{r} - \bald{u} \, t),
}
where $J^\nu$ is the source term, $dE/dt$ is the rate of energy loss (to be discussed in the next section, see Figure \ref{elossfigure}) and $u^\nu = (1,\bald{u})$ with $\bald{u}$ being the velocity of the parton shower. This form of the source neglects the broadening of the parton shower due to finite emission angles but is adequate for the purpose of conserving energy and momentum, which is the focus here.  Additionally, the energy and momentum deposited by this source term are equivalent, provided $|\bald{u}|\rightarrow 1$ which is adopted in what follows.  Thus the momentum balance should recover the amount of energy deposited by the parton shower.

In order to obtain the final particle distribution for the CF freezeout spectrum it is necessary to evaluate both the energy and momentum density perturbation generated by the parton showers.  To illustrate how these are obtained, I focus on the energy density perturbation, $\delta \epsilon$, which has the momentum space representation in linearized hydrodynamics \cite{Neufeld:2010tz}
\eqf{\label{eps}
\delta\epsilon ({\mathbf r},t) = \int \frac{d^4 k}{(2\pi)^4} e^{- i k \cdot x} 
\frac{i k J_L(k)  + J^0(k)(i \omega -  \Gamma_s \bald{k}^2)}
{\omega^2 -  c_s^2 \bald{k}^2 + i \Gamma_s \omega \bald{k}^2}. 
}
In the above equation, $c_s$ denotes the speed of sound and $\Gamma_s = \frac{4 \eta }{3 s T}$ is the sound 
attenuation length with $\eta/s$ being the shear viscosity to entropy density ratio in the medium.  
Also, the source vector has been divided into transverse and longitudinal parts: 
${\mathbf J} = \hat{\mathbf k} J_L + {\mathbf J}_T$.  

For the time-dependent source term of equation (\ref{simplesource}) it is convenient to write the momentum space form as
\eqf{\label{timesource}
J^\nu(k) &= \int_0^{t_s} d t' \frac{dE}{dt}(t')\,u^\nu\,e^{-i \bald{k}\cdot\bald{u} t' + i \omega t'}
}
where the $t'$ integration runs over the interval during which the parton shower is depositing energy.  In equation (\ref{timesource}) this interval is from $t = 0$ to $t = t_s$, at which point the parton shower stops interacting with the medium.  So explicitly
\eqf{\label{finnyform}
\delta\epsilon ({\mathbf r},t) &= \int \frac{d^4 k}{(2\pi)^4} \int_0^{t_s} d t' \frac{dE}{dt}(t')\,u^\nu \\
&\times e^{- i\omega (t-t') + i \bald{k} \cdot(\bald{r} - \bald{u} t')} \frac{i \bald{u}\cdot\bald{k}  + (i \omega -  \Gamma_s \bald{k}^2)}
{\omega^2 -  c_s^2 \bald{k}^2 + i \Gamma_s \omega \bald{k}^2}. 
}
where $t > t_s$.  If the parton shower continues to deposit energy indefinitely, then the upper limit of the $t'$ integration would simply be $t$.

In this paper, $t$ in equation (\ref{finnyform}) represents the moment of freezeout, and $t_s$ is taken to be $8$ fm (for reasons discussed in the following section).  The difference between these two times, $t - t_s$, is here referred to as the 'gap' time.  The gap time is a freely adjustable parameter which will be varied to see how it affects the CF freezeout spectrum.

\subsubsection{Calculating the Parton Shower Energy Deposition Rate}

The parton shower energy deposition rate is an essential quantity in evaluating the medium response and in turn the momentum deposition spectrum, and in the following paragraphs I will discuss the approach used here to evaluate it.  A fast parton evolves into an in-medium parton shower through medium-induced radiative processes. These processes are here constrained by the Gyulassy-Levai-Vitev (GLV) radiative energy loss formalism \cite{Vitev:2005he}.  I make use of the bremsstrahlung spectra for jets averaged over the collision geometry in central PbPb reactions at the LHC.  These spectra have been previously employed to discuss jet and particle production in heavy-ion collisions at the highest $\sqrt{s_{NN}}$ \cite{lhcrad}, and more recently to consider next to leading order jet production at $\sqrt{s_{NN}}$ = 2.76 GeV \cite{He:2011pd}.  

As a rough approximation of the scenario observed in dijet measurements, I will consider 120 GeV dijets produced in $\sqrt{s_{NN}}$ = 2.76 GeV heavy-ion collisions.  Using results based on the analysis of \cite{He:2011pd} one finds for these parameters that the probability that the primary partons initiating the dijets are two gluons is about 37$\%$, the probability for two quarks is about 21$\%$, and the probability for a quark and a gluon is about 42$\%$.  Thus almost 80$\%$ of dijet events contain at least one primary gluon jet.  Further, because the average squared color charge of the gluon is larger than that of the quark, gluons on average lose more energy to the medium than quarks.  This creates a bias such that secondary jets in dijet events are most often initiated by gluons.

With the assumption that the secondary jet deposits the bulk of the energy, it makes sense to focus on 120 GeV gluon initiated parton showers.  However, corrections for the primary jet and ratio of secondary jets initiated by quarks versus gluons will be estimated in the next section, so I also include the quark initiated parton shower energy deposition for later reference.  The energy and momentum deposited by the primary jet should not change the qualitative results of this section, other than to reduce the net momentum deposited along the axis of the secondary jet.  The medium induced gluon emission is probabilistic and depends on the parton species. The most likely number for a 120 GeV gluon propagating in a QGP produced in a $\sqrt{s_{NN}}$ = 2.76 GeV heavy-ion colilsion is 14 with average energy $\omega_g \approx 4.8$~GeV and average emission angle $\theta \approx 0.7$.  For a 120 GeV quark, the most likely number changes to 6 with average energy $\omega_g \approx 6$~GeV.  The parton shower energy deposition is of a collisional nature.  The rate of this energy deposition is here taken from the recent results on parton shower energy loss by Neufeld and Vitev \cite{ev2}, which was an extension of the work done in \cite{operatorsource}.  In this work, the finite time effects associated with the formation of the gluon in the medium were included as well as the quantum color interference between primary parton and radiated gluon.  

In order to have a consistent treatment, one should account for the softening of the parton shower due to the collisional energy losses.  Although this effect was not accounted for in the work in \cite{ev2}, it can be implemented in a phenomenological way by using the rate of energy transfer from each parton in the shower as a differential equation to evolve the energy of that parton.  The rate of energy transfer from each parton here depends logarithmically on the energy through an ultraviolet cutoff characteristic of collisional energy loss and enters as $dE/dt  \sim  m_D^2 \ln (1.6\sqrt{E T}/m_D)$ where $T$ is the temperature of the medium, $E$ is the energy of the parton, $m_D$ is the Debye mass, and the numerical factor of 1.6 in the logarithm was evaluated in \cite{Thomas:1991ea} for number of active quark flavors, $N_F$, equal zero.

\begin{figure}
\centerline{
\includegraphics[width = 0.75\linewidth]{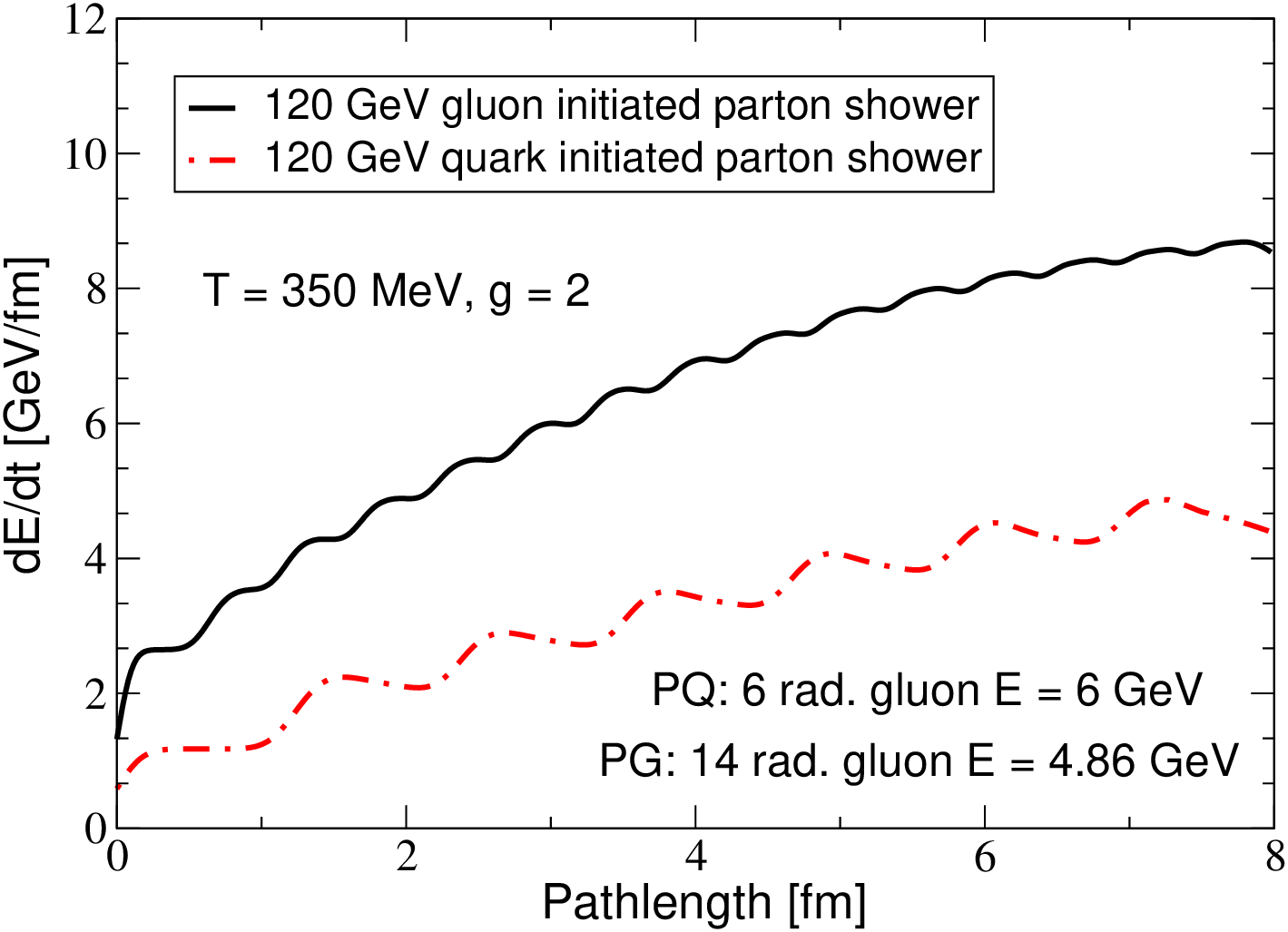}
}
\caption{(Color online) Energy deposition rates from parton showers initiated by a 120 GeV gluon (PG) and 120 GeV quark (PQ) to the underlying medium as a function of pathlength.  The results are obtained using the formalism given by \cite{ev2} - which was an extension of the work done in \cite{operatorsource}.  Details are discussed in the text.
}
\label{elossfigure}
\end{figure}

As in \cite{ev2}, the quantum color interference effects between the primary and radiated partons are taken to be associated with the rate of energy transfer from the radiated gluon.  For the purposes of extracting the energy deposition the gluon emission points are uniformly distributed along the parton propagation path in a medium of length 8 fm (this pathlength is based on the analysis of \cite{He:2011pd}).  With these details in place, the result for the energy loss rate as a function of parton shower pathlength is shown in Figure \ref{elossfigure} for the parameters: $g = 2$, $T = 0.35$ GeV and $m_D = g T$, which are based on the average values obtained for LHC collision energies in a Bjorken expanding plasma, and one has $N_F = 0$ for a gluon-dominated medium.  The total integrated energy deposition shown in Figure \ref{elossfigure} amounts to about 51 GeV for the primary gluon and about 25.5 GeV for the primary quark.  As mentioned, I will focus on gluon initiated parton showers in the rest of this section, so it will be necessary to verify 51 GeV is recovered in the momentum balance performed in this section.  The result of the gluon energy deposition in Figure \ref{elossfigure} is used in the source term of equation (\ref{simplesource}).

The details of the evaluation of the parton shower energy deposition rate and implementation into linearized hydrodynamics have intentionally been made brief here in order to keep the presentation concise.  The important point of this section is to show the qualitative features of the final spectrum of parton shower momentum deposition and how that compares to the distribution of the underlying medium.  The interested reader is directed to the references listed above for more detailed explanations.

\subsubsection{Results on the Momentum Deposition Spectrum}

With the ingredients discussed above, it is now possible to obtain the momentum deposition spectrum for a parton shower propagating in the $\hat{x}$ direction.  I will examine the parton shower momentum deposition spectrum as a function of two parameters.  The first is the shear viscosity to entropy density ratio.  The detailed structure of the medium response to the parton shower energy deposition is highly sensitive to the shear viscosity to entropy density ratio.  A medium with small shear viscosity, such as believed to be formed in experimentally accessible heavy-ion collisions \cite{Romatschke:2007mq}, is characterized by a well defined disturbance in the presence of a parton shower.  As one increases the shear viscosity, the disturbance generated by the parton shower becomes less well defined and more smeared.  However, no matter how the details of the medium response change as the shear viscosity is varied, the total energy and momentum are conserved.  As is shown below, this serves to constrain the form of the final momentum deposition spectrum such that it does not change much as the shear viscosity is varied.

The second parameter I will consider is the time between the end of the parton shower energy deposition and freezeout.  This 'gap' time (recall equation (\ref{finnyform}) and the following paragraph) provides a simple parameter with which to simulate the expansion of the parton shower due to finite emission angle and broadening resulting from scattering with the medium.  The point souce $\delta$ function approximation adopted above is convenient for implementation into hydrodynamics but a realistic parton shower broadens as a function of time.  The gap time provides a convenient way to mimick this broadening and see how it influences the final spectrum.

\begin{figure}
\centerline{
\includegraphics[width = 0.75\linewidth]{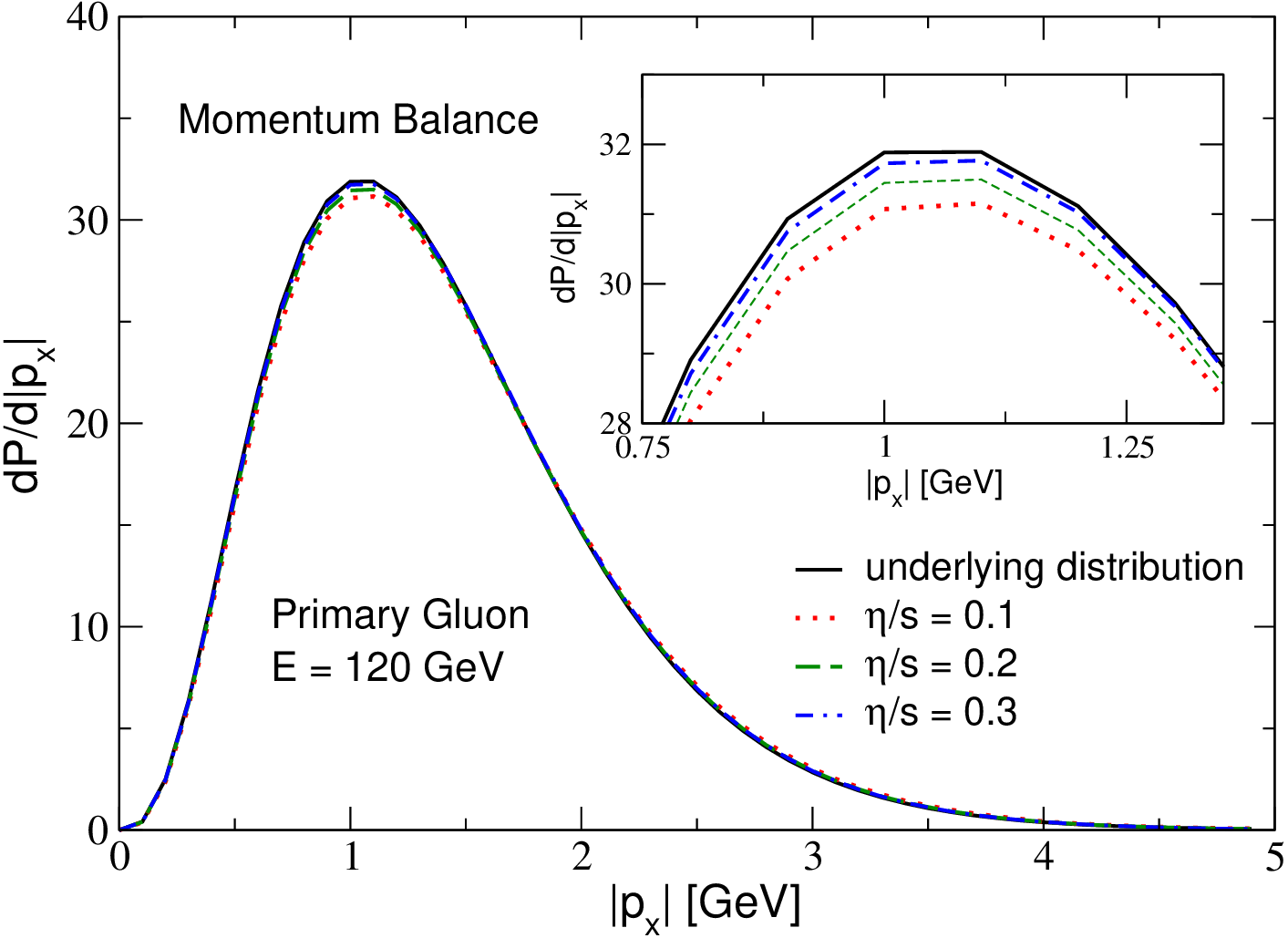}
}
\caption{(Color online) The final freezeout momentum deposition spectrum generated by a parton shower is largely independent of the details of how the transferred energy and momentum are redistributed - in this case, as determined by the shear viscosity.   The momentum deposition spectrum is defined in equation (\ref{momentum}) and the underlying distribution in equation (\ref{underlie}).
}
\label{balance1}
\end{figure}

\begin{figure}
\centerline{
\includegraphics[width = 0.75\linewidth]{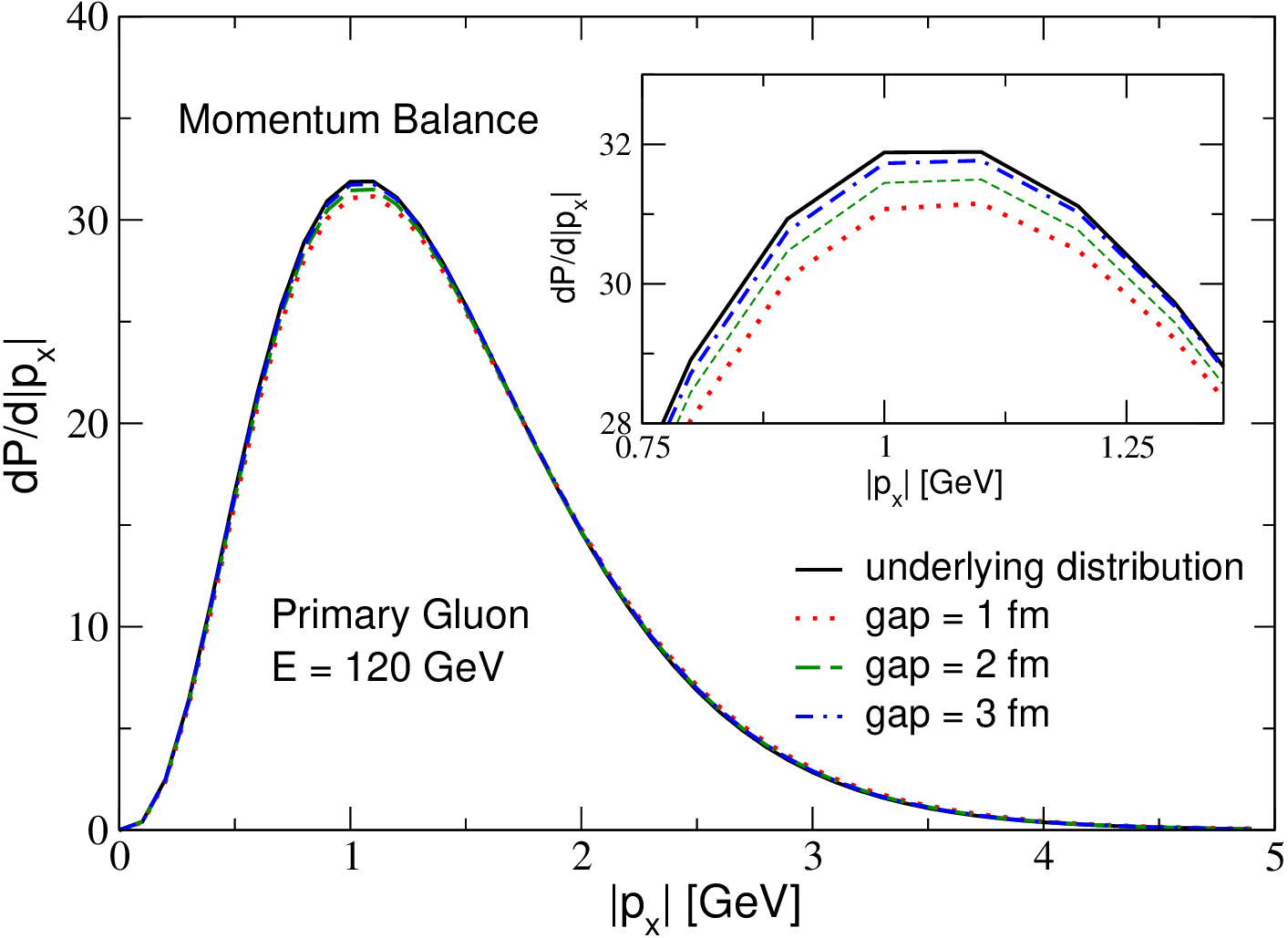}
}
\caption{(Color online) As in Figure \ref{balance1}, but in this case the gap time is varied, which roughly characterizes how broad the parton shower is.  The shear viscosity to entropy density ratio is kept at 0.2 in all curves.}
\label{balance2}
\end{figure}

I now present the results on the differential momentum deposition spectrum for different shear viscosity to entropy density ratio, along with the underlying medium distribution, in Figure \ref{balance1}.  The momentum deposition spectrum is defined in equation (\ref{momentum}), and the underlying distribution in equation (\ref{underlie}).  For all the curves, the calculation was performed with a gap time of 2 fm.  The underlying distribution has been normalized to 50 GeV in order to match up with the momentum balance curves.  Additionally, the background medium is at rest in this scenario.  

The three momentum deposition curves in Figure \ref{balance1} each integrate to about 50 GeV, slightly less than the 51 GeV obtained from Figure \ref{elossfigure}.  This  discrepancy results from numerical limitations in the hydrodynamic solutions and subsequent freezeout, however the difference is rather small.  Notice also that the momentum balance curves are plotted as $dP/d|p_x|$, so that both positive and negative contributions are contained.  This is necessary to recover all of the deposited momentum.  

Figure \ref{balance1} shows that the final spectrum associated with the parton shower energy deposition, or the momentum deposition spectrum, is largely independent of the details of how the energy and momentum are redistributed - in this case, as determined by the shear viscosity.   The three curves are almost indistinguishable at a first glance, and follow nicely the behavior obtained from linearizing the CF equation above (recall this is different than linearizing the hydrodynamic equations), even though that linearization is not adopted in the results presented here.  As mentioned, the results obtained from the hydrodynamic solution are kept in their full functional form when evaluating the final spectrum.  It is clear from the inset that as the shear viscosity is made larger, the momentum deposition spectrum moves closer to the underlying distribution.  This makes sense intuitively, since larger viscosity tends to push the disturbance generated by the parton shower to softer scales and into the regime where the CF equation can be linearized.

I next present the results on the differential momentum deposition spectrum for different gap times, along with the underlying medium distribution, in Figure \ref{balance2}.   For all the curves, the calculation was performed with a shear viscosity to entropy density ratio of 0.2.  One finds a similar result as in Figure \ref{balance1}, that the final spectrum depends only modestly on the gap time, which loosely represents how broad the parton shower is.  Again the three curves integrate to about 50 GeV, and the underlying distribution has been normalized accordingly.  From the inset it is clear that the momentum deposition spectrum moves closer to the underlying distribution as the gap time is increased.  This occurs for the same reason as when the shear viscosity is increased in Figure \ref{balance1}.  In both cases the disturbance generated by the parton shower move further into the regime where the CF equation can be linearized at the time of freezeout.

When combining the results of Figures \ref{balance1} and \ref{balance2} a clear picture emerges that the results of the parton shower momentum deposition spectrum are largely independent of the details of how deposited energy is redistributed in the medium.  As argued for in the Introduction, this is a non-trivial result considering how significantly these details change when varying the shear viscosity and gap time.  However, the constraint that momentum be conserved along with the general form of the CF freezeout formula seems to provide a limitation on the amount that the momentum deposition spectrum is modified.  Any deviations in the momentum deposition spectrum as compared to the underlying distribution tend to be in the form of a slight blueshift.

\begin{figure*}
\centerline{
\includegraphics[width = 0.47\linewidth]{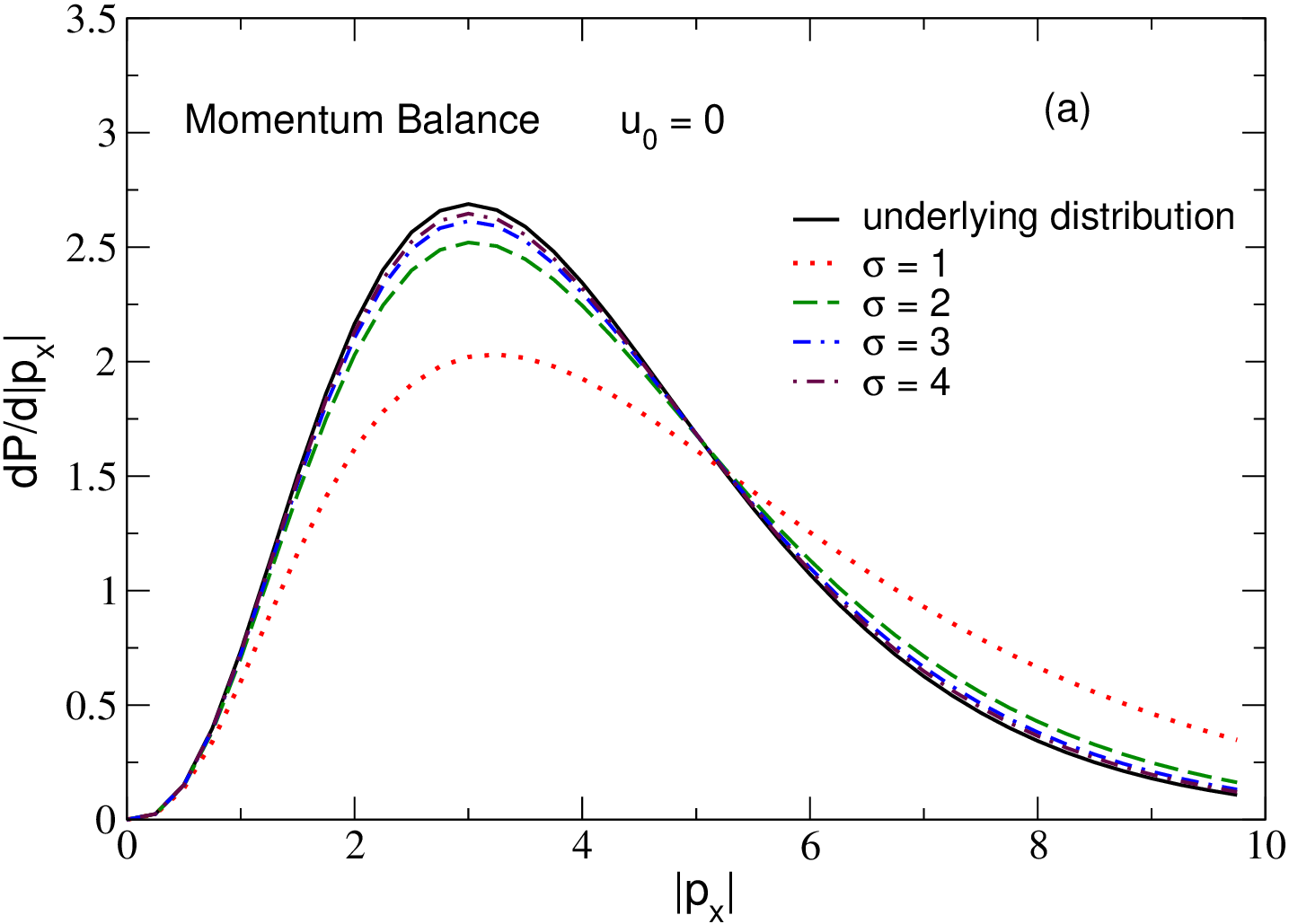}\hskip0.05\linewidth
\includegraphics[width = 0.45\linewidth]{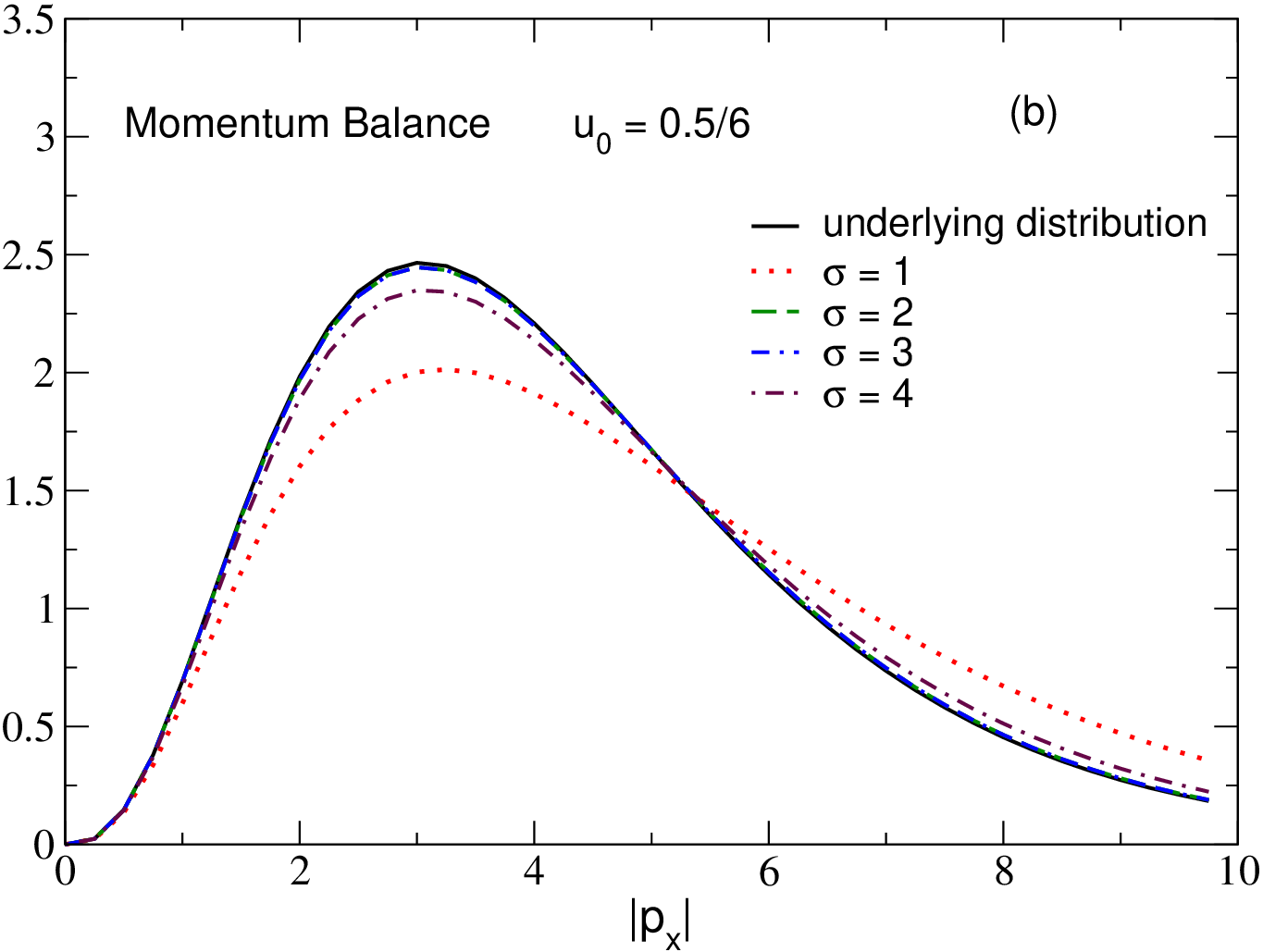}
}
\caption{(Color online) The freezeout distribution for the one dimensional flow with a perturbation as described in equations (\ref{trickflow}) and (\ref{perturbedflow}) - along with the underlying distribution.  The parameter $\sigma$ characterizes how localized the perturbation is.  As long as the perturbation is not too localized, the result depends only insensitively on the value of $\sigma$.  In each case, the momentum deposition spectrum has some amount of blueshift relative to the underlying distribution.
}
\label{contrived1}
\end{figure*}

The relative independence of the parton shower momentum deposition spectrum on the details of how the energy and momentum are deposited and redistributed in the medium is a potentially useful feature, particularly in the case that one can somehow relate it to the distribution of the underlying medium.  For instance, in the case considered above, the parton shower momentum deposition spectrum follows almost exactly the distribution of the underlying medium.  Thus, if one had results on jet momentum balance and wanted to know how much of that momentum actually was related to parton shower energy deposition - once again emphasizing that not all of the 'lost' energy is deposited in the medium - it would be possible to obtain a nice estimate using the underlying distribution.  Of course, the ultimate goal is to obtain this estimate from the jet momentum balance measurements performed in heavy-ion collisions.  In that case, there are strong underlying flow fields, so it is not obvious how well the jet momentum deposition spectrum will match the underlying distribution.  Before leaving this section, I will consider a simple example to get a feel for how the above results might change when considering a medium with underlying flow.

\subsection{A Contrived Example with Underlying Flow Fields}

In this subsection, I consider a simple contrived example involving a perturbation on top of an underlying flow field in order to get a qualitative sense for how the results presented above change in the presence of flow.  Start with a system with flow velocity in the $x$ direction at the moment of freezeout given by: 
\eqf{\label{trickflow}
\bald{u} = \baldh{x}\,u_0\,x ,
}
where $u_0$ is some free parameter, and the medium runs from $x = \pm 6$ with a cross sectional area of 1 in the $y$ and $z$ directions - units are arbitrary here.  Equation (\ref{trickflow}) is symmetric in $x$, so the underlying flow does not contribute to the momentum balance.  Now consider adding a perturbation to the flow such that the total momentum contained in the perturbation is small compared to the total energy of the unperturbed system.  To constrain the form of the perturbation, I note that for a system of the volume described above obeying Boltzmann statistics with uniform temperature $T_0 = 1$ the total energy contained is
\eqf{
E_T = \frac{36}{\pi^2}.
}
If the perturbed flow is written as $\delta u$ then the total momentum contained in the perturbation will be
\eqf{
\frac{3}{\pi^2} \int d^3 x \,\delta u.
}
This remains a perturbation on the system for instance, if $\int d^3 x \,\delta u = 1$.  With this motivation, the perturbation is given the explicit form
\eqf{\label{perturbedflow}
\bald{u} \rightarrow \baldh{x}\,u_0\,x + \baldh{x}\,\frac{\Theta(x - \sigma)\Theta(\sigma - x) }{2\sigma},
}
where the range of allowed $\sigma$ depends on the choice of $u_0$ (the total velocity should remain less than 1) but can never exceed 6 because of the volume under consideration.

In this example, $\sigma$ plays a role similar to shear viscosity or the gap time.  As $\sigma$ is increased, the perturbation smears out over the volume, and as $\sigma$ is decreased the perturbation becomes more localized and well defined.  However, no matter what the value of $\sigma$ is, the total momentum contained in the perturbation remains constant.  Now I present results on the spectrum of the momentum perturbation and the underlying distribution, similar to what was done above for the parton shower energy deposition.  I first choose $u_0 = 0$ and vary $\sigma$ from 1 to 4 by one.  The result is shown in Figure \ref{contrived1}a.  The scenario for $u_0 = 0$ is rather similar to the case of the parton shower energy deposition in linearized hydrodynamics and one naively expects there to be a similar type matching to the underlying distribution, which is indeed the case for $\sigma = 2,3,4$.  One starts to see deviation for the case of $\sigma = 1$, which shows a moderate blueshifting of the momentum deposition spectrum.  From equation (\ref{perturbedflow}) it is clear that $\sigma = 1$ has a magnitude of 0.5 and is rather localized.  It seems that if the perturbed flow becomes too localized, the momentum balance spectrum begins to blueshift away from the underlying distribution.

Now I present the result for $u_0 = 0.5/6$ in Figure \ref{contrived1}b.  This corresponds to an underlying flow field that varies from 0 at the center to 0.5 at the edge of the volume.  Interestingly, the qualitative features of the result do not change much when including flow in the example considered here.  Again, whatever deviation there is from the underlying spectrum results in a slight to moderate blueshift of the momentum deposition spectrum.

I here summarize the main observations of this section:

\begin{itemize}
\item  The momentum deposition spectrum generated by a parton shower (defined in equation (\ref{momentum})) in linearized hydrodynamics is relatively independent of the details of how the energy is redistributed in the medium.  Furthermore, the spectrum matches quite closely to the shape of the underlying medium distribution (defined in equation (\ref{underlie})) as seen in Figures \ref{balance1} and \ref{balance2}.
\item When considering the simple example of one dimensional flow with a perturbation, the previous conclusion still holds, except in the case when the deposited momentum is very localized in the medium, as seen in Figure \ref{contrived1}.
\item Whatever deviations there are in the shape of the momentum deposition spectrum as compared to the underlying medium distribution tend to be in the form of a blueshift.
\end{itemize}

In the next section I will apply these conclusions to the jet momentum balance measurements performed by CMS to obtain an estimate on the parton shower energy deposition in heavy-ion collisions.

\section{Extracting the Energy Deposition from Jet Momentum Balance}\label{III}

Assuming the results of the previous section hold in the case of a heavy-ion collision, it should be possible to make a reasonable estimate on parton shower energy deposition using experimental results.  In this section I examine the CMS results on dijet momentum balance \cite{LHCIII} and compare to results on the bulk particle spectra presented by ALICE \cite{alice}.  In this way the amount of parton shower momentum deposited in these events is estimated, including corrections made for the primary jet and ratios of quark to gluon initiated jets.

\begin{figure}
\centerline{
\includegraphics[width = 0.75\linewidth]{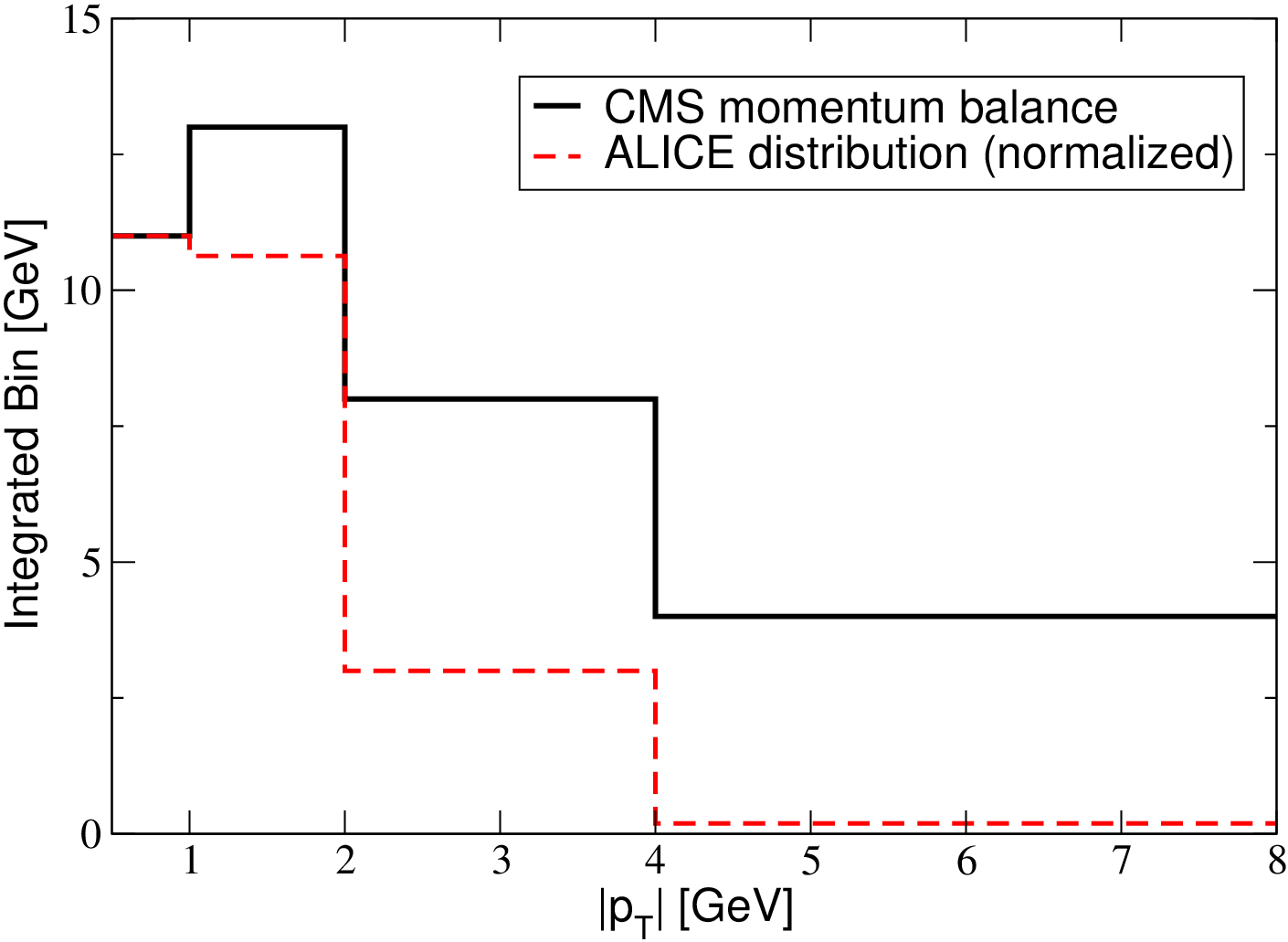}
}
\caption{(Color online) Using the appropriately normalized $p_T$ distribution of primary charged particles at mid-rapidity for central collisions \cite{alice}, the momentum balance for dijet events with asymmetry of 0.4 as measured by CMS \cite{LHCIII} is matched to the underlying distribution - as defined by Equation (\ref{experiment}).  This type of matching provides a lower limit on the parton shower energy deposition.  The comparison suggests that about 25 GeV of the 36 GeV momentum projected onto the secondary jet axis is part of the parton shower energy deposition.
}
\label{cmsalice}
\end{figure}

I consider the dijet events with $A_j = 0.4$ as shown in Figure 14 of reference \cite{LHCIII}, and make the assumption that only tracks with momentum projection less than 8 GeV are part of the parton shower energy deposition, from which there is a total of about 36 GeV available.  As discussed in the Introduction, not all of the 36 GeV momentum projected onto the secondary jet axis is part of the parton shower energy deposition - only some portion of the medium induced energy loss is transferred to the medium.  The expectation is that the momentum balance should should contain a thermal contribution from the energy deposition and a non-thermal contribution from whatever is not transferred to the medium.

To estimate what fraction of the 36 GeV is from the energy deposition, I follow the conclusions of the previous section and match the momentum balance distribution to the distribution of the underlying medium.  Using the appropriately normalized $p_T$ distribution of primary charged particles at mid-rapidity for central collisions \cite{alice} for the underlying distribution - as defined by Equation (\ref{experiment}) - the momentum balance is matched, as in Figure \ref{cmsalice}.  The matching is done by assuming the two spectrums are identical in the smallest momentum bin of 0.5-1 GeV.  Using the smallest momentum bin for the matching makes sense because one expects the thermal medium to dominate the spectrum in the low $p_T$ region.  This type of matching likely provides a lower limit on the parton shower energy deposition, since according to the last section the momentum deposition spectrum tends to blueshift relative to the underlying distribution.  The comparison in Figure \ref{cmsalice} is consistent with the expectation that the dijet momentum balance contains both a thermal and non-thermal part.  From Figure \ref{cmsalice} it is estimated that about 25 GeV of the 36 GeV momentum projected onto the secondary jet axis is part of the parton shower energy deposition.

\begin{figure}
\centerline{
\includegraphics[width = 0.85\linewidth]{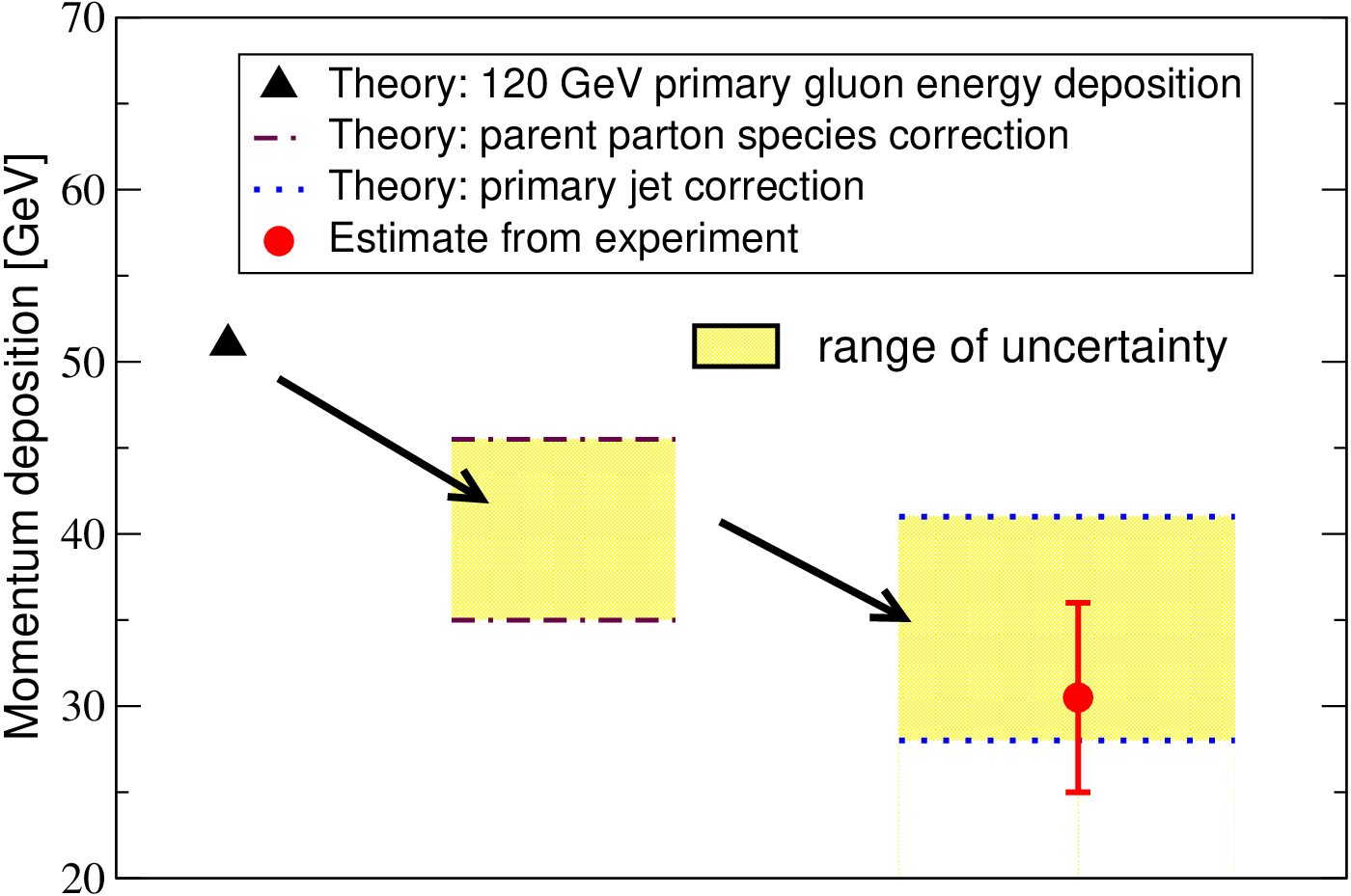}
}
\caption{(Color online) To compare single parton shower energy and momentum deposition to what can be extracted directly from dijet measurements requires some model input, including corrections for the primary jet and the ratio of secondary jets initiated by quarks versus gluons.  These corrections are discussed in the text and shown in the Figure along with the uncertainty.  The experimental estimate is obtained by matching momentum balance results to the underlying distribution.
}
\label{corrections}
\end{figure}

This number is smaller than the 51 GeV calculated in the previous section for a 120 GeV gluon traveling about 8 fm in medium.  However, to estimate single parton shower energy and momentum deposition from dijet measurements some model dependence must enter.  One needs to estimate the corrections for the primary jet and the ratio of secondary jets initiated by quarks versus gluons.  These corrections are estimated in the discussion that follows, and are graphically demonstrated in Figure \ref{corrections} for the reader's convenience.

The first correction to consider is that not all of the secondary jets are initiated by gluons.  Using the precise percentages quoted in section \ref{II}, at least 21$\%$ of the secondary jets are initiated by quarks.  Further, for the sake of argument, assume that from the 42$\%$ of dijets coming from a quark and a gluon the secondary jets are all initiated by gluons (quarks).  In that case, the energy deposition of secondary jets averages to 45.5 (35) GeV, as seen in the first band of Figure \ref{corrections}.  The band represents the possible range one can obtain from these two extreme scenarios.

Another source of uncertainty is in ignoring the contribution of the primary jet.  Estimates based on the analysis of \cite{He:2011pd} suggest the primary jet travels about 2 fm in medium.  The energy and momentum deposited by the primary jet should not change the qualitative results of the previous section, but it will reduce the net momentum deposited along the axis of the secondary jet.  Using the result of Figure \ref{elossfigure}, there is about 7.3 GeV deposited by a primary gluon in 2 fm of pathlength, and 3.2 GeV for a primary quark.  Again assuming that from the 42$\%$ of dijets coming from a quark and a gluon the primary jets are all initiated by quarks (gluons), the energy deposition of primary jets averages to 4.7 (6.4) GeV.  This energy deposition serves to reduce the net momentum deposited by the dijet pair.   Consistently keeping track of the previous two paragraphs yields a range of 28-41 GeV predicted net momentum deposited by the dijet pair, which is illustrated by the second band of Figure \ref{corrections}.

The final source of uncertainty is from the tendency for the momentum balance spectrum to be blueshifted with respect to the underlying distribution.  For this reason, the 25 GeV estimate above is likely a lower limit as long as the assumption that the two spectrums are identical in the smallest momentum bin holds.  At the other extreme, one could consider that all of the contribution less than 8 GeV projected along the secondary jet axis is part of the parton shower energy deposition.  This amount, which totals 36 GeV, serves as an upper limit and is shown along with the 25 GeV lower limit as the 'experimental estimate' in Figure \ref{corrections}.  As one can see from the Figure, once these corrections are implemented the theory prediction and the estimate from experiment are compatible with each other.

One can also work backwards from the 25 GeV estimate made above to obtain an estimate for the amount of energy deposited by the secondary jet.  For example, the result of Figure \ref{elossfigure} shows there is about 7.3 GeV deposited by a primary gluon in 2 fm of pathlength, and 3.2 GeV for a primary quark.  These numbers are roughly 1/8 of the total amount deposited in 8 fm pathlength.  Using this ratio and ignoring any uncertainty coming from parent parton species, it is found that the secondary jet deposits about 29 GeV.  If one instead uses the upper limit estimate that all of the contribution less than 8 GeV projected along the secondary jet axis is part of the parton shower energy deposition, one finds there is about 41 GeV deposited by the secondary jet.  So I estimate that in dijet events with asymmetry of 0.4, the secondary jet deposits about 30-40 GeV of energy.

Clearly any uncertainties regarding pathlength and parent parton ratios tend to make the extraction of the secondary jet energy deposition less precise.  Although the net momentum deposition from the dijet system can be estimated in a model independent way by matching the distribution of momentum associated with the jet momentum balance
to the thermal distribution of the underlying medium, making corrections for the primary jet and parent parton ratios requires some model input.  For instance, if the relative pathlengths change significantly from the estimates quoted above, the extraction of the secondary jet energy deposition also changes.  

To more precisely constrain the parton shower energy deposition, momentum balance measurements for jets tagged with electroweak bosons are ideal.  In this way, one significantly reduces the uncertainty due to pathlength and fraction of quark versus gluon initiated jets and can obtain a truly model independent estimate of single parton shower energy deposition.

\section{Summary and Conclusions}\label{IIII}

In this paper I have shown that jet momentum balance measurements provide an opportunity to quantify parton shower energy deposition in heavy-ion collisions.  This is true because the momentum deposition spectrum generated by a parton shower (equation (\ref{momentum})) is relatively independent of the details of how the energy is redistributed in the medium, which is largely determined by transport coefficients such as shear viscosity.  This was shown explicitly in section \ref{II} for the momentum deposition spectrum generated by a parton shower in linearized hydrodynamics.  Furthermore, the spectrum matches quite closely to the shape of the underlying medium distribution (defined in equation (\ref{underlie})) as seen in Figures \ref{balance1} and \ref{balance2}.  Whatever deviations there are in the shape of the momentum deposition spectrum as compared to the underlying medium distribution tend to be in the form of a blueshift.

These observations were applied in section \ref{III} to the CMS results on dijet momentum balance and results on the bulk particle spectra presented by ALICE.  From the comparison I obtained a lower limit estimate that 25 GeV of net momentum is deposited by dijets in events with asymmetry of 0.4.  This estimate was compared to theoretical predictions presented in Figure \ref{elossfigure}.  After making appropriate corrections for parent parton species and the primary jet, the theory prediction is compatible with the estimate from experiment, as seen in Figure \ref{corrections}.

In conclusion, the main results from this paper are:

\begin{itemize}
\item  For the cases considered here, the momentum deposition freezeout spectrum generated by a parton shower is relatively independent of the details of how the energy is redistributed in the medium.  Furthermore, the spectrum matches quite closely to the shape of the underlying medium freezeout distribution as seen in Figures \ref{balance1} and \ref{balance2}. Whatever deviations there are in the shape of the momentum deposition spectrum as compared to the underlying medium distribution tend to be in the form of a blueshift.
\item Applying the previous point to the case of dijet events measured by CMS with $A_j = 0.4$ and matching to the underlying distribution using results on the bulk particle spectra from ALICE, a lower limit estimate is made that 25 GeV of net momentum is deposited in the direction of the secondary jet axis, with an upper limit of 36 GeV.  From these numbers, it is estimated that the secondary jet deposits about 30-40 GeV of energy and momentum in the underlying medium.  This is compatible with the theoretical predictions shown in Figure \ref{elossfigure}.
\item The previous two points also apply to momentum balance for jets tagged with electroweak bosons.  The advantage of using tagged jets is that the uncertainty due to pathlength and fraction of quark versus gluon initiated jets is minimized and one can obtain a truly model independent estimate of single parton shower energy deposition.
\end{itemize}

\small{{\it Acknowledgments}: I wish to thank Ivan Vitev for a critical reading and helpful remarks, as well as Ulrich Heinz for pointing out an error in an earlier version.  I also thank Yuncun He, Ivan Vitev and Ben-Wei Zhang for providing numerical values for the medium induced radiation and dijet cross section analysis.  This work was supported in part by the US Department of Energy, Office of Science, under Contract No. DE-AC52-06NA25396.}

\end{document}